\documentclass[epj,twocolumn]{webofc}
\usepackage[varg]{txfonts}   

\woctitle{Powders \& Grains 2017}

\usepackage{enumitem}

\begin{document}

\title{A local view on the role of friction and shape}
\author{\firstname{Matthias} \lastname{Schr\"oter}\inst{1}\fnsep\thanks{\email{matthias.schroeter@fau.de}}}
\date{\today}

\institute{Institute for Multiscale Simulation, Friedrich-Alexander-University (FAU), 
91052 Erlangen, Germany.}

\abstract{Leibniz said "Naturam cognosci per analogiam": nature is understood by making analogies. 
This statement describes a seminal epistemological principle. 
But one has to be aware of its limitations: quantum mechanics for example at 
some point had to push Bohr's model of the atom aside to make progress.  
This article claims that the physics of granular packings has to move beyond the 
analogy of frictionless spheres, towards local models of contact formation.
}

\maketitle

On earth solid assemblies of granular particles are by far the most frequent phase of granular matter;
we encounter granular packings everywhere from our kitchen cabinet to civil engineering textbooks.
In order to make their handling, transport, and storage more efficient, we strive for a theory 
that predicts their mechanical properties, such as shear and bulk modulus or yield stress, 
starting from a few state variables only. Efforts to develop such a theory often start by modeling 
granular packings as an assembly of frictionless spheres. This is a rather unsuitable starting point, for a 
number of reasons:

\begin{enumerate}[topsep=0pt,itemsep=-0.8ex,partopsep=1ex,parsep=1ex]
\item All granular particles are frictional.

\item Frictional particles have lower isostatic numbers than frictionless particles.

\item Granular physics happens at volume fractions inaccessible to frictionless particles.

\item The volume fraction of soft particles can be changed by compression. The volume fraction of frictional 
particles is changed by changing their geometry.
 
\item Friction is one reason for history dependence in granular systems.

\item Real world granular media are rarely spherical. Shape adds complexity, e.g. to history dependence. 
\end{enumerate}

These six theses are also the outline for the following sections. They are intended to provoke discussions 
with a sizeable subgroup of the theoretically or numerically working scientists. 
Many experimentalists, applied scientist, and engineers might find them, at least in part, well-known. 
For simplicity, we will discuss in the following only monodisperse spheres; except for section \ref{sec:shape}.

\section{All granular particles are frictional.}
\label{sec:gran_frict}
\begin{figure}[t]
\begin{center}
  \includegraphics[width=0.42\textwidth]{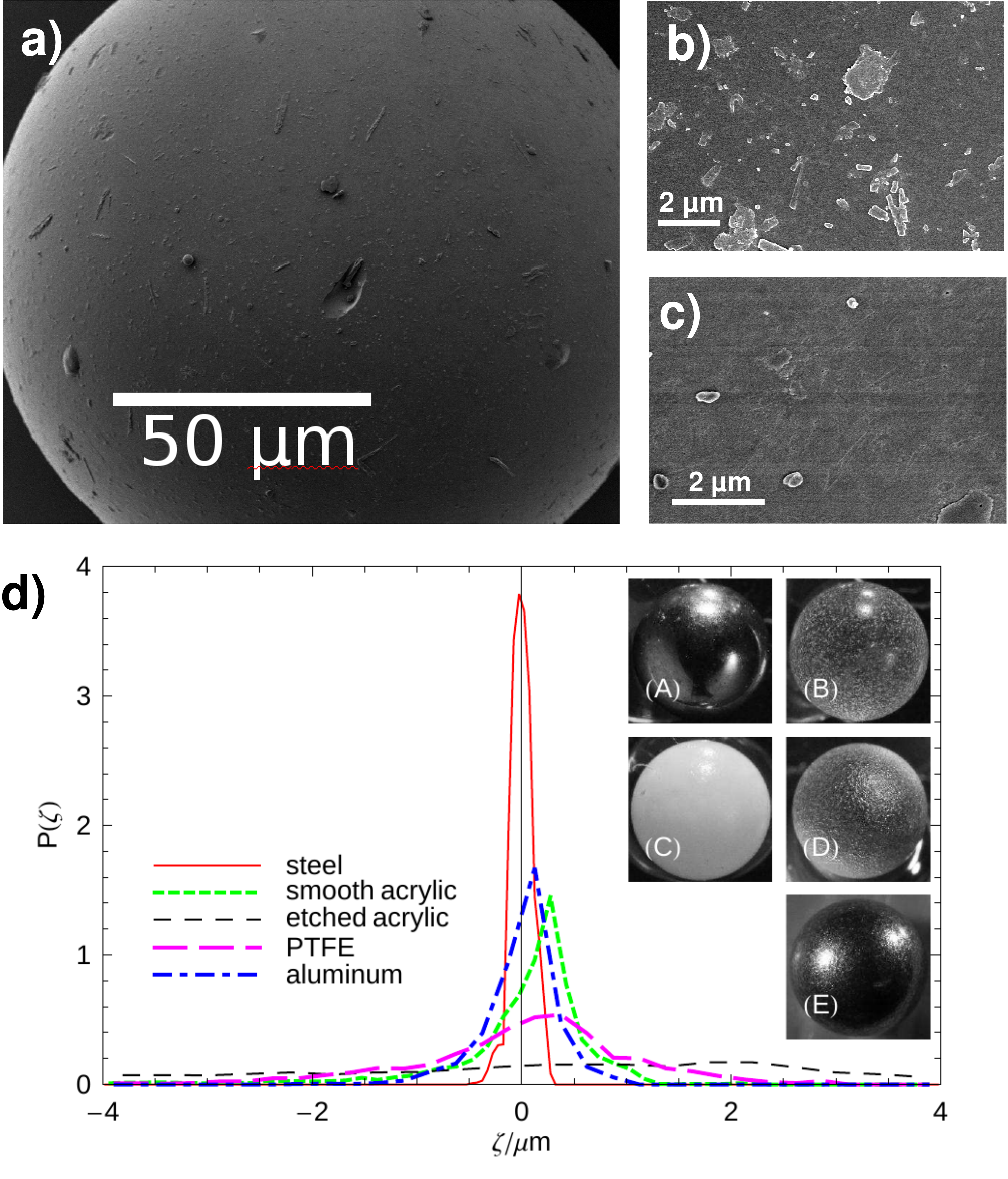}
      \caption{\small 
      Granular particles are rough particles. 
      {\bf a)} Scanning Electron microscope (SEM) image  of a factory-new soda-lime glass bead.
      Image by courtesy of Karina Sand.
      {\bf b)} SEM image of the asperities on the surface of a new soda-lime glass bead. 
      {\bf c)} After 30900 flow pulses in a water fluidized bed, abrasion has removed many of the surface asperities,
      resulting in a measurable difference in packing properties.
      Images b) and c) are from \cite{schroeter:05}.
      {\bf d)} Histograms of surface roughness $\xi$ of spherical particles measured with a profilometer. 
      The inset shows images of the corresponding particles: (A) steel, (B) smooth acrylic, (C) PTFE, 
      (D) etched acrylic, (E) aluminum. Adapted with permission from \cite{farrell:10}.
      }
    \label{fig:surface_sem}
\end{center}
\end{figure}

Contrary to other particulate systems, such as foams or emulsions, 
the constituents of  granular media are solid particles. This implies that their 
surface is geometrically rough, cf. figure \ref{fig:surface_sem}.  
If two particles get into contact, their 
surface asperities will  interlock, allowing for the existence of tangential forces 
at the contact \cite{bowden:73}.  
In the context of granular packings, friction is sufficiently well described by the
Amontons-Coulomb law:
\begin{equation}
  F_t \le \mu F_n
\label{eq:friction}
\end{equation}
$F_n$ and $F_t$ are the normal and tangential components of the contact force, and 
$\mu$ is the static coefficient of friction. 

All granular media consist of 
frictional particles. Even hydrogel spheres, which consist of up to 99.5\% water, have
a friction coefficient of $\approx$ 0.01 \cite{dijksman:13,dijksman:17}. Moreover, while it is possible 
to relax all tangential forces in a packing by vibrating it at small amplitudes and high frequencies
\cite{gao:09}, this will also compactify the packing to values falling into the range 
of frictionless packings, thereby bypassing the interesting granular physics as discussed in section \ref{sec:frict_pack}.

\begin{figure}[t]
  \includegraphics[width=0.5\textwidth]{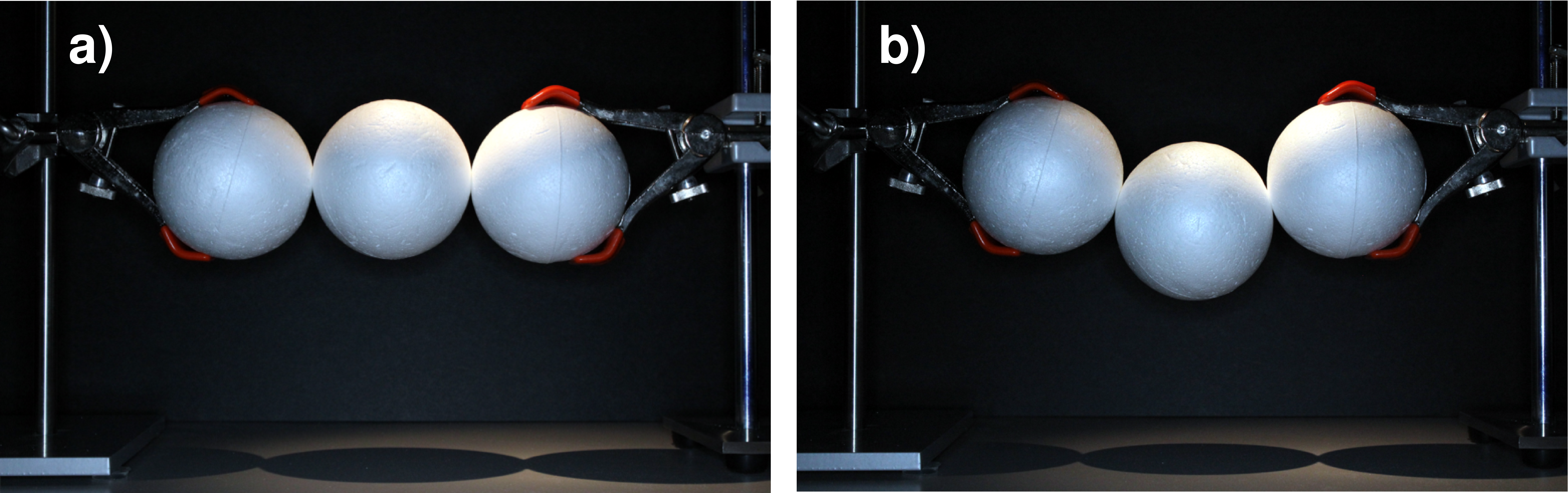}
      \caption{\small 
       Friction increases the number of mechanically stable configurations.
      In both images the center Styrofoam sphere is only hold in place by the tangential forces at 
      the contacts.
      Neither arrangement would be possible with frictionless spheres.
      In fact, under gravity the only mechanically stable configuration of three frictionless spheres is the perfect vertical alignment.  
      }
    \label{fig:three_spheres_in_a_row}
\end{figure}

In loose granular systems, such as e.g.~granular gases, the dynamics is more controlled by 
collisions than contacts. Friction changes the way particles exchange momentum during collisions, but this 
seems to be often only a higher order perturbation.
Granular packings on the other hand  consist of enduring contacts, 
here the existence of friction changes the physical picture completely:

 First, the presence of tangential forces provides additional ways to satisfy the force and torque balance,
which will be discussed more quantitatively in section \ref{sec:frict_low_iso}. This leads to a massive 
increase of the number of mechanically stable states; figure \ref{fig:three_spheres_in_a_row} gives a simple example.
Most granular packings studied in nature or experiment are looser than the loosest packing that can be created without friction.
The consequences  of this will be discussed in sections \ref{sec:frict_pack} and \ref{sec:local_theories}.

\begin{figure}[t]
\begin{center}
  \includegraphics[width=0.31\textwidth]{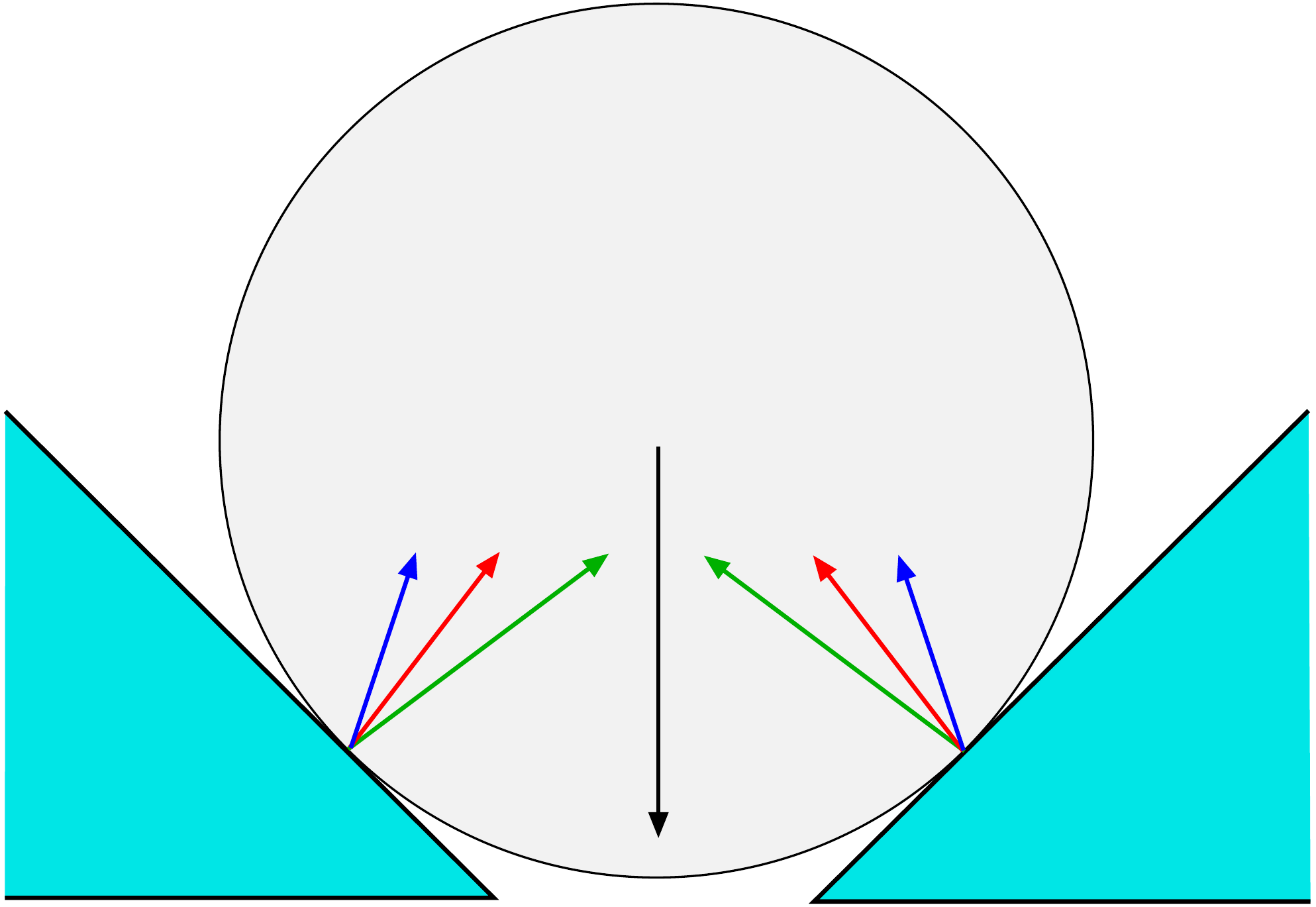}
      \caption{\small The contact forces of a sphere in a wedge depend on the
        preparation history. Each of the blue, red, or green pairs of contact forces can balance the weight
        of the sphere (black arrow). Which one is realized depends how the sphere was placed.
      }
    \label{fig:sphere_in_wedge}
\end{center}
\end{figure}

Secondly, equation \ref{eq:friction} is an inequality. For a given normal force at a contact it allows for 
a whole range in tangential force, as shown in figure \ref{fig:sphere_in_wedge}. 
The actual tangential force will be a consequence of how the contact was formed. 
This property is one of the reasons for the so called history-dependence of granular matter, which will be 
discussed further in section \ref{sec:hist_dep}.

\section{Frictional particles have lower isostatic numbers than frictionless particles.}
\label{sec:frict_low_iso}
For any granular packing to be solid, the  average  number of contacts  a particle forms with its neighbors $Z$ 
has to be at least so large that all degrees of freedom (DOF) of the  particle can be fixed.
This minimal value, the so called isostatic contact number $Z_{iso}$, does depend on the dimension, shape,
and most importantly friction of the particles considered. 

In the absence of friction, the rotational DOF of a perfect sphere are not relevant and only
the three translational DOF  have to be blocked by the contacts. At each contact there exists 
one normal force, which is however ``shared'' between the two particles which means that each 
contact blocks on average only 0.5 DOF per particle. A packing of frictionless spheres needs therefore
to have at least $Z_{iso}^0 = 6$ contacts to be mechanically stable. 

If we assume $\mu = \infty$, then  each contact has 3 independent force components (one normal and two tangential), which 
fix 1.5 constraints per particle. On the other hand, we now also have to  consider  the rotational DOF which results in 6 DOF
per particle. The isostatic number in the presence of infinite friction $Z_{iso}^{\mu}$ is therefore 4.

The inequality $Z_{iso}^0 > Z_{iso}^{\mu}$ holds for all particle shapes and in 2 and 3 dimensions \cite{van_hecke:10}.
One consequence is the massive increase  in  the  number of mechanically stable packings of frictional particles,
as discussed in the next section. Another consequence is that granular packings are typically {\it hyperstatic}
i.e.~their actual contact number is larger than the minimum number needed for stability: $Z > Z_{iso}^{\mu}$. 
From this follows that for a given spatial configuration of particles, there exists a multitude 
of possible force networks that will all satisfy the boundary conditions of the system \cite{tighe:10}. 
This property is intimately connected to the history dependent behavior discussed in section \ref{sec:hist_dep}.

\begin{figure}[t]
\begin{center}
        \includegraphics[width=0.42\textwidth]{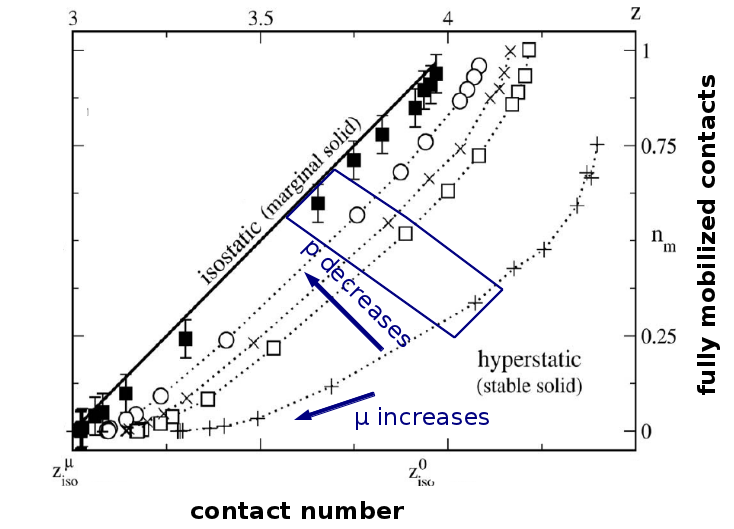}
    \caption{ \small
      Most granular packing are also hyperstatic  when fully mobilized contacts are taken into account.
      In disc packings both $Z$ (x-axis) and the number of fully mobilized contacts (y-axis)
      changes with pressure and $\mu$. However, only in the limit of vanishing pressure the system
      will approach isostaticity (solid line).
      Adapted with permission from \cite{shundyak:07}.
      }
    \label{fig:contact_numbers}
\end{center}
\end{figure}

There is a possible caveat regarding hyperstaticity. In real granular media $\mu$ is 
finite and the contacts might have tangential forces which are 
exactly at the Coulomb threshold, so called fully mobilized contacts. As this type of contact will block 
only 1 DOF, the constraint counting argument has  to be modified and the effective $Z_{iso}^{\mu}$ becomes larger. 
However, a number of numerical studies \cite{silbert:02,zhang:05,shundyak:07,henkes:10} have shown that the number 
of fully mobilized contacts is not sufficient to regain isostaticity in any other situation than when preparing  a pressure free 
packing very slowly, cf figure \ref{fig:contact_numbers}.
Which is also the recipe to approach the limit of Random Loose Packing,
the loosest packing possible (discussed in section \ref{sec:frict_range}).

\section{Granular physics happens at volume fractions inaccessible to frictionless particles.}
\label{sec:frict_pack}
The consequences of the lower isostatic contact number of frictional particles are best discussed using the
the concept of the configurational entropy $S_{conf}$ of the packings.
$S_{conf}$ was first introduced by Sam Edwards \cite{edwards:89,mehta:89}, it is proportional 
to the logarithm of the the number of mechanically stable packing configurations that fit in a given volume
and support given boundary conditions. As we are interested in the thermodynamic limit,
we will discuss here $S_{conf}$ as a function of the global volume fraction $\phi_g$.
More specifically, we are interested in a comparison of the upper and lower bounds of $\phi_g$ between which
 $S_{conf}$ becomes non-zero for both frictional and frictionless systems.

\begin{figure}[t]
\begin{center}
  \includegraphics[width=0.44\textwidth]{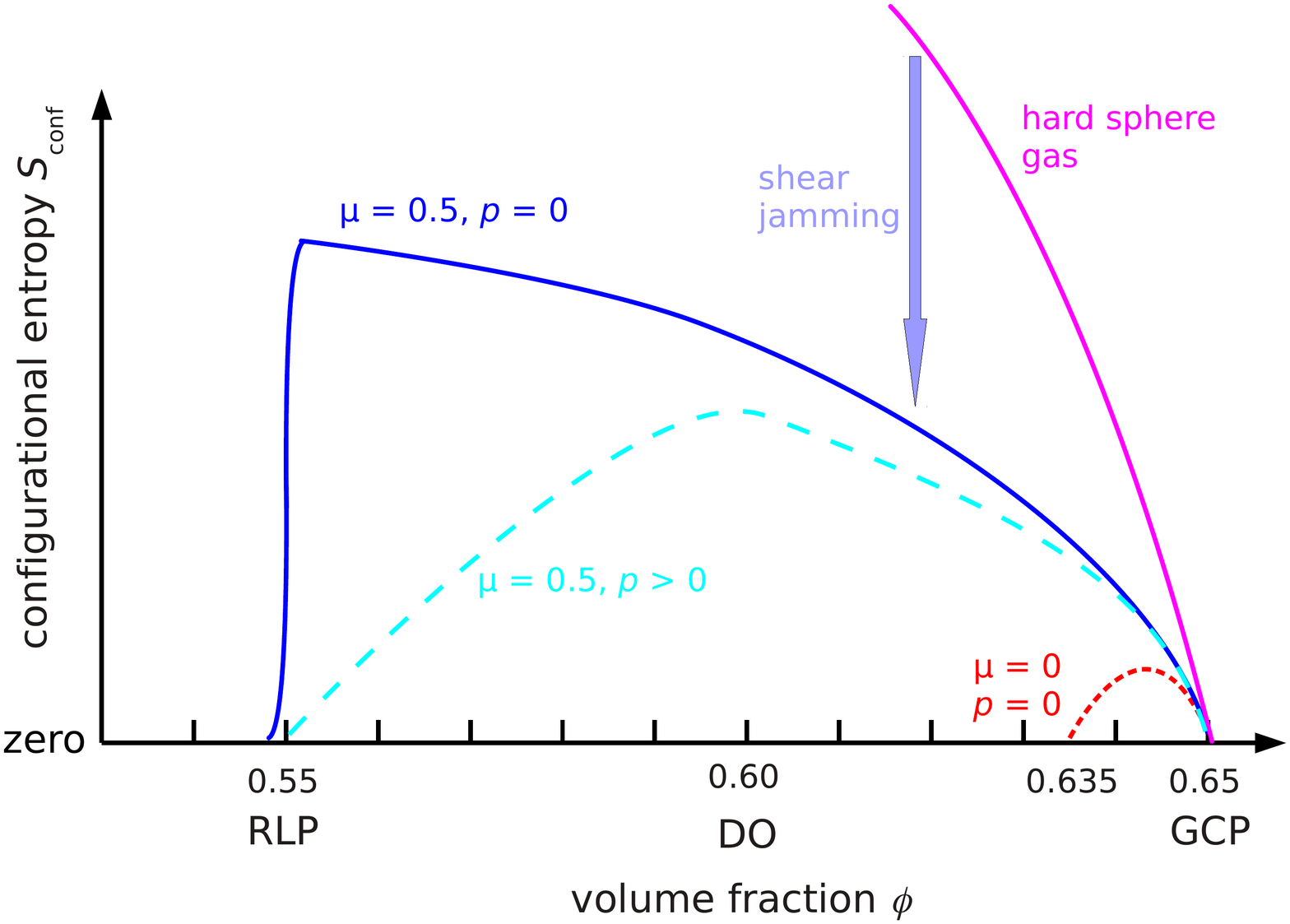}
      \caption{\small A schematic how
        the configurational entropy $S_{conf}$ of a sphere packing depends on the friction coefficient $\mu$  and the confining 
        pressure $p$. The solid magenta line indicates the  configurational entropy of an amorphous hard sphere gas  
        i.e.~a "packing" with non overlapping particles but no requirements on the mechanical stability or number of contacts. 
        The solid blue and dashed cyan line represent packings with an approximate real world value of $\mu$ and the dotted red line
        corresponds to frictionless particles.
        While the points with $S_{conf} = 0$ are well supported by experiments and simulations, the actual shape of the different 
        curves is speculative.
      }
    \label{fig:conf_entropy}
\end{center}
\end{figure}

The main results of this discussion are summarized in figure \ref{fig:conf_entropy}. 
But a word of caution is necessary:
The $\phi_g$ values of the upper and lower bounds are well supported by the numerical 
and experimental work discussed below. But the functional form of   $S_{conf}$ connecting these boundaries
is  speculative and only supported by the heuristic arguments given below.

 Moreover, we are only interested in packings that exist in a thermodynamic sense. This excludes both packings 
crystallizing in FCC and HCP configurations at volume fractions above $\phi_{g} \approx 0.65$
\cite{anikeenko:07,radin:08,jin:10,kapfer:12,francois:13,baranau_random_close:14,baranau_jamming:14}
\footnote{Readers familiar with equilibrated hard sphere systems might expect crystallization to occur in the range 
$\phi_g \approx 0.494 - 0.61$. However, these systems are driven by the entropy increase due to newly 
gained vibrational DOF. These DOF do not exist in athermal granular packings where all particles 
are permanently in contact.}  
and the ``tunneled crystal'' packings at $\phi_g$ = 0.49 \cite{torquato:07}.
Neither of these two configurations are extensive, i.e.~their number does not grow exponentially 
with the number of particles in the system. Which means that in the thermodynamic limit their entropy is zero.

An important upper bound on  $S_{conf}$ is the configurational entropy  $S_{HS}$ of an amorphous {\it hard sphere gas} where the only 
remaining condition for a valid configuration is that particles do not overlap. Mechanical stability and 
consequentially $Z$ do not matter. In  figure \ref{fig:conf_entropy} this boundary is indicated by a solid magenta line. 
The pressure of a hard sphere gas diverges at the so called Glass Close Packing (GCP) point 
with $\phi_{GCP} \approx 0.65$ \cite{parisi:10,baranau_random_close:14,baranau_jamming:14}. 
From which follows that the system runs out of non-overlapping configurations and $S_{HS}$ goes to zero.

\subsection{Frictionless sphere packings: $ 0.635< \phi_g <  0.65$   } 
As $S_{HS}$ is an upper limit for any sphere packing, $S_{conf}$ of a frictionless packing 
also needs to go to zero at $\phi_{GCP} \approx 0.65$. 
For any smaller value of $\phi_g$, $S_{conf}$ has to be smaller than $S_{HS}$ 
because we now additionally require an isostatic number of contacts. 
In fact, the set of mechanically stable configurations should be of measure zero compared to hard sphere gas: 
there are infinite more possibilities of two spheres to be not in contact compared to the one configuration 
where they are. Luckily, we know that  $S_{conf}$  is still extensive. 
This was shown for soft frictionless disk and sphere systems of different sizes by dividing the 
total accessible phase space volume by that of an average basin of 
attraction \cite{asenjo:14,martiniani:16}.

The total range of $S_{conf} \ne 0$ is shown as a red dotted line  in figure \ref{fig:conf_entropy}.
There is still some debate \cite{ohern:03, pica_ciamarra:10, van_hecke:10,baranau_jamming:14}
if the onset of mechanical stability happens for frictionless spheres
at the so called Jamming point of $\phi_{J} \approx 0.64$ or slightly below at  $\phi_{g} \approx 0.635$. 
However, it is known that the actual 
volume fraction of a packing of uncompressed  frictionless spheres will depend on the preparation history.
For an extended discussion  and further references see \cite{kumar:16}.  

\subsection{ Frictional sphere packings: $ 0.55 < \phi_g <  0.65$   } 
\label{sec:frict_range}
All mechanically stable configuration of frictionless particles will stay valid if we allow for additional tangential forces.
Therefore  $S_{conf}$ of frictional packings will always be larger than its frictionless counterpart. 
Because  $S_{HS}$ is also an upper bound to frictional systems, GCP will still be the upper limit for uncompressed packings.

The lower boundary,  commonly referred to as Random Loose Packing (RLP), is however considerably lower than in frictionless systems:
As mentioned in the last section, the inequality $Z_{iso}^0 > Z_{iso}^{\mu}$ holds for all particle shapes and in 2 and 3 dimensions. 
Moreover, as $Z$ can be generically expected to decrease monotonically with decreasing $\phi_g$ (i.e.~larger average separation between particles), 
the onset of  mechanical stability  will happen at a  lower volume fraction for frictional particles. 
The actual value of $\phi_{\text{RLP}}$  does depend on pressure \cite{onoda:90,jerkins:08}
and the friction coefficient\cite{jerkins:08,farrell:10}: the higher $\mu$ the smaller is 
$\phi_{\text{RLP}}$. 
For the experimentally common values of $\mu \approx  0.5$ and vanishing pressure, $\phi_{\text{RLP}}$ approaches
0.55 \cite{onoda:90,jerkins:08,song:08,farrell:10,silbert:10,delaney:11}.

In figure \ref{fig:conf_entropy} the solid blue and the dashed cyan line  represent  $S_{conf}$ of sphere packing with a finite
value of $\mu$ and either zero or finite confining pressure $p$. Besides the points with  $S_{conf} = 0$, the 
shape of the curves is speculative as there are few analytical or experimental results on  how $S_{conf}$ depends on $\phi_g$. 
What we do understand is that for any given value of $\phi_g$,  $S_{conf}$ will grow monotonically with $\mu$  because 
allowing larger tangential forces will never destabilize any existing packing, but allow for new, additional  configurations \cite{srebro:03}. 
Moreover, we can use the Widom insertion method in combination with experimental or numerical packings to obtain an upper bound on 
 $S_{conf}$ \cite{baranau:16}. However, a lower bound would be more helpful. Finally, under certain 
additional assumptions, $S_{conf}$ can be computed from the volume fluctuations of a repeatedly driven
granular packing \cite{briscoe:08,mcnamara:09,zhao:14}. However, the results obtained this way 
do not agree with each other.

Another way of assessing the shape of $S_{conf}(\phi_g)$ is to consider experimental preparation protocols and to use the 
additional assumption that the state  the system will end up in is the most likely one: 
the one with the highest value of
 $S_{conf}$ under the given circumstances. For example, packing prepared by slow sedimentation in an almost density matched 
fluid (i.e.~in the limit $p \rightarrow 0$) will always end up at $\phi_{\text{RLP}}$ \cite{onoda:90,jerkins:08,farrell:10}.
Which indicates that $S_{conf}$ has a maximum at RLP for $p = 0$, cf. the blue line in figure \ref{fig:conf_entropy}.  

Without density matching (i.e.~at a finite static pressure) and with an increased sedimentation speed 
(meaning that the settling particles will transfer more momentum on the already existing packing),
the most likely packing fraction moves up to $\phi \approx 0.6$ \cite{schroeter:05}. 
The cyan dashed line in figure \ref{fig:conf_entropy} represents the idea that this becomes the new maximum in  $S_{conf}$.
Which can be rationalized by assuming that configurations with less excess contacts compared to an isostatic packing
are more likely to not posses a force network capable of supporting the increased stress at the boundaries. 
$\phi \approx 0.6$ is incidentally also the value for the onset of dilatancy, 
which is discussed in the next subsection. 

Finally, getting the system to compactify to values of $\phi_g$ above 0.6 requires repeated driving under 
confining gravitational pressure, either by flow pulses \cite{schroeter:05} or mechanical taps \cite{nowak:98,ribiere:07}.  
This indicates that these states become more and more unlikely which agrees with the idea that  $S_{conf}$ goes to zero for
$\phi_g$ approaching GCP.

But the main point of this section is untouched by this discussion of the shape of  $S_{conf}$:
The range of volume fractions of mechanically stable packings is {\it seven times larger} for frictional particles
than for frictionless particles.

\subsection{Dilatancy in frictional packings}
Most of the interesting physics of granular packings happens at intermediate values of $\phi_g$ (i.e.~between RLP and GCP); 
these volume fractions are  inaccessible to frictionless packings.  A good example is dilatancy:
If a dense granular packing is sheared (at a finite hydrostatic pressure), it will expand \cite{andreotti:13}.
However, dilatancy does not occur in sufficiently loose samples;  those will instead compactify.

Now if dense packings expand during shear and loose packings collapse, 
there will be an intermediate density, usually termed Dilatancy Onset (DO) or critical state, where the volume fraction 
$\phi_{DO}$ stays constant during shear.  Due to effects like shear banding the exact determination of 
$\phi_{DO} $ is not straightforward, but most experiments point to $\phi_{DO} \approx 0.6$ 
for frictional spheres at comparatively small confining pressures  
\cite{schroeter:07,kabla:09,gravish:10,umbanhowar:10,metayer:11,mueggenburg:12}

Dilatancy is closely related to a phenomenon labeled shear-jamming \cite{bi:11,ren:13,vinutha:16}. 
For $\phi_g > \phi_{DO}$ one can start from a
hard sphere gas configuration (experiments are done in horizontal two-dimensional system and therefore 
effectively in the absence of gravity), shear it at a constant volume and arrive at a mechanically stable configuration.
This protocol is indicated as a blue arrow in figure  \ref{fig:conf_entropy}. Below $\phi_{DO}$ this is not possible.

One way of interpreting $\phi_{DO}$ is to see it as a "natural attractor" for all sheared systems starting at other volume fractions.
This interpretation agrees well with the idea of $S_{conf}$ being maximal at $\phi_{DO} \approx 0.6$.

\section{The volume fraction of frictional particles is controlled by their geometry.}
\label{sec:local_theories}

The exact value of $\phi_g$ of a frictionless packing at zero pressure does depend on the
preparation history \cite{baranau_random_close:14,baranau_jamming:14,kumar:16,luding:16}. 
However, the scaling laws  of these packings are normally studied by preparing pressure free packings
with a given protocol and then increasing $\phi_g$  by compressing this packing \cite{ohern:03, van_hecke:10}.

With respect to the contact number $Z$ this amounts to a study of the pair correlation function, or more precisely
the slope of the right shoulder of the first peak which describes the close-by particles which will form contacts when 
compressed. This slope leads to an equation for $Z$:  
\begin{equation}
  \label{eq:jamming}
  Z(\phi_g)= Z_{iso}^0 + c (\phi_g - \phi_{iso})^{0.5}
\end{equation}
Here $\phi_{iso}$ is the volume fraction of the uncompressed, isostatic packing and
the constant $c$ depends on the dimension and polydispersity of the system.
For compressed frictionless packings such as emulsions and foams \cite{katgert:10}, equation \ref{eq:jamming} is indeed a good description.

However, real world granular particles are normally not very squishy; 
they change their volume fraction by isobarically changing their packing geometry not by compression.  
We all intuitively know this from kitchen physics: 
if we want to fill more grains into a storage container we do not compress   
them with a piston, but we tap the container a couple of times on the counter top to change its packing structure.

For a more quantitative example lets compare two glass spheres  
(Young's modulus =  70 GPa, diameter = 250 $\mu$m) which are either uncompressed at the upper surface or compressed below
a 1 m high column of other glass spheres. Using Hertz law we can derive that this increase in pressure 
will deform the sphere by approximately 10 nm at each contact. This
deformation is an order of magnitude smaller than the vertical surface 
roughness of typical glass spheres\cite{utermann:11}.  Assuming that the 
sphere is compressed symmetrically, this corresponds to a change in volume fraction of
$7 \times10^{-5}$  compared to the uncompressed sphere. This illustrates that the
large range of $ 0.55 < \phi_g <  0.65$ available to frictional sphere packings 
can not be explored by compression.

Please note that granular experiments can be performed in a way to test frictionless models. 
E.g.~the compression of frictional but sufficiently soft photoelastic discs (with a Young's modulus of 4MPa \cite{majmudar:07}) 
can be used to verify equation \ref{eq:jamming} \cite{majmudar:07} or study glassy behavior \cite{coulais:14}; 
provided that the tangential forces are relaxed by additional tapping or vibration.    
These experiments do however not prove that frictionless models describe generic frictional particles.

\subsection{Friction with your neighbors? Think locally!}
Because  $Z$ and $\phi_g$  are in frictional packings 
not simultaneously controlled by the globally defined parameter pressure, the 
idea  expressed in  equation \ref{eq:jamming} of a function  $Z(\phi_g)$  runs into an epistemological problem. 
Contacts are formed at the scale of individual particles and their neighbors.
At this scale the global $\phi_g$ is not only undefined; due to local volume correlations \cite{lechenault:06,zhao:12}
it would even be impossible for a particle 
scale demon to compute $\phi_g$ by averaging over the volume of the neighboring particles.  

What is needed for the theoretical description of frictional particles is an ansatz which 
explains $Z$  using only locally defined (i.e.~on a particle level) parameters \cite{aste:06,baule:13,baule:14,schaller:15}.
The most important \cite{schaller:15} of these local parameters  is the 
local volume fraction $\phi_l$ which describes by how much free volume an individual particle is surrounded.
$\phi_l$ is computed by dividing the volume of the particle by the volume of its Voronoi cell (a tessellation method assigning
all points in space to the closest particle). 
However, for a complete local description more parameters such as the shape of the Voronoi cells \cite{schroeder-turk:10}
or the fabric anisotropy \cite{radjai:09} are needed.

\begin{figure}[t]
\begin{center}
  \includegraphics[width=0.40\textwidth]{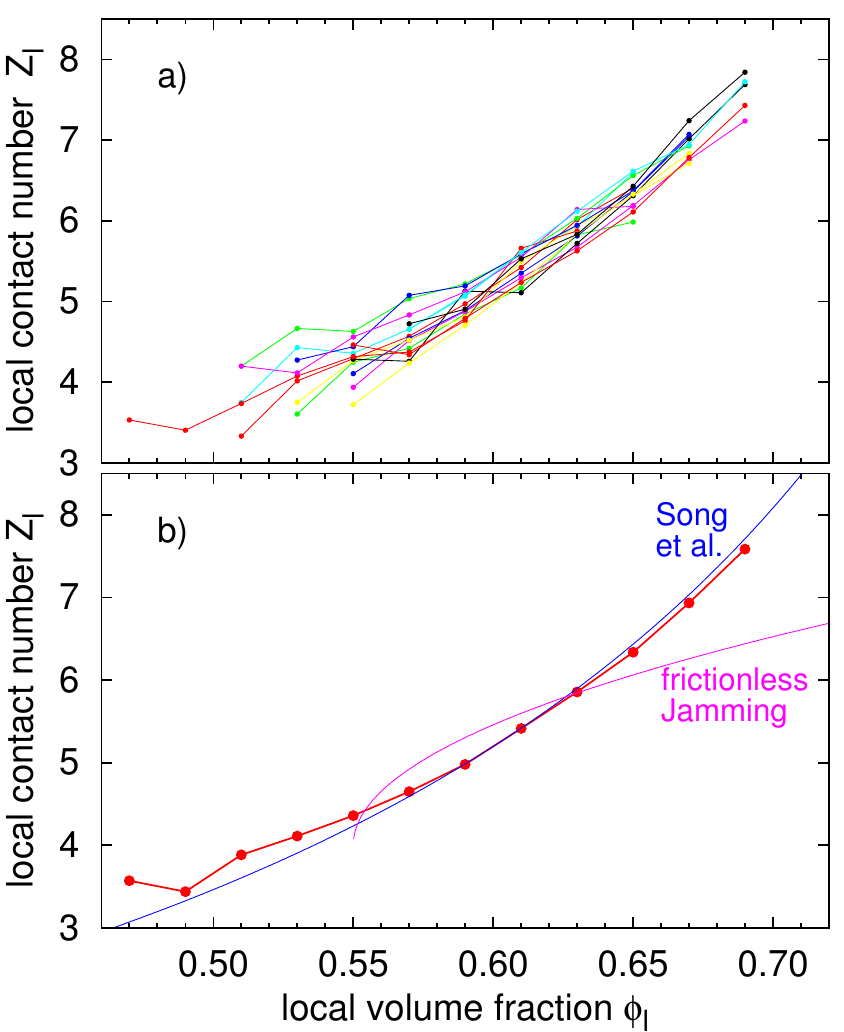}
      \caption{\small Understanding contact numbers in sphere packings requires a local approach. 
        {\bf a)}  The average local contact number $Z_l$ of individual spheres, 
        measured by X-ray tomography and averaged in local volume fraction  $\phi_l$ bins of size 0.02.
        Data corresponds to 15 different sphere packings with global volume fractions $\phi_g$
        in the range from 0.56 to 0.625.
        Within experimental scatter, $Z_l$ depends only on $\phi_l$, not on $\phi_g$.
        {\bf b)} The red dots correspond to a bin-wise average of all data shown in panel a. The local mean field
        theory by Song {\it et al.}~(eq.~\ref{eq:makse}, no fit parameter) provides a fair description of the data.
        This can not be said about the local interpretation of the scaling law for frictionless, compressed spheres
        (eq.~\ref{eq:jamming}, one fit parameter).
        From \cite{schaller:15}.
      }
    \label{fig:local_Z}
\end{center}
\end{figure}

Figure \ref{fig:local_Z}  substantiates this claim for the necessity of local theories. 
Panel a shows that the number of contacts $Z_l$ an individual particle will form does only depend on its own  $\phi_l$, 
not the $\phi_g$ value of the packing it resides in. Figure  \ref{fig:local_Z} b) demonstrates that
$Z_l$ can be well explained by the local theory presented in \cite{song:08}, which predicts:
\begin{equation}
  \label{eq:makse}
  Z= \frac{2 \sqrt{3} \; \phi_l}{1- \phi_l}
\end{equation}

A local reinterpretation of equation  \ref{eq:jamming} for frictional systems 
($\phi_g$ becomes $\phi_l$, $Z_{iso}^0$ becomes $Z_{iso}^{\mu}$, 
$\phi_{iso}$ becomes $\phi_{RLP}$) fails to describe the experimental data.

\section{Friction is one reason for history dependence in granular systems.}
\label{sec:hist_dep}
\begin{figure}[t]
\begin{center}
  \includegraphics[width=0.40\textwidth]{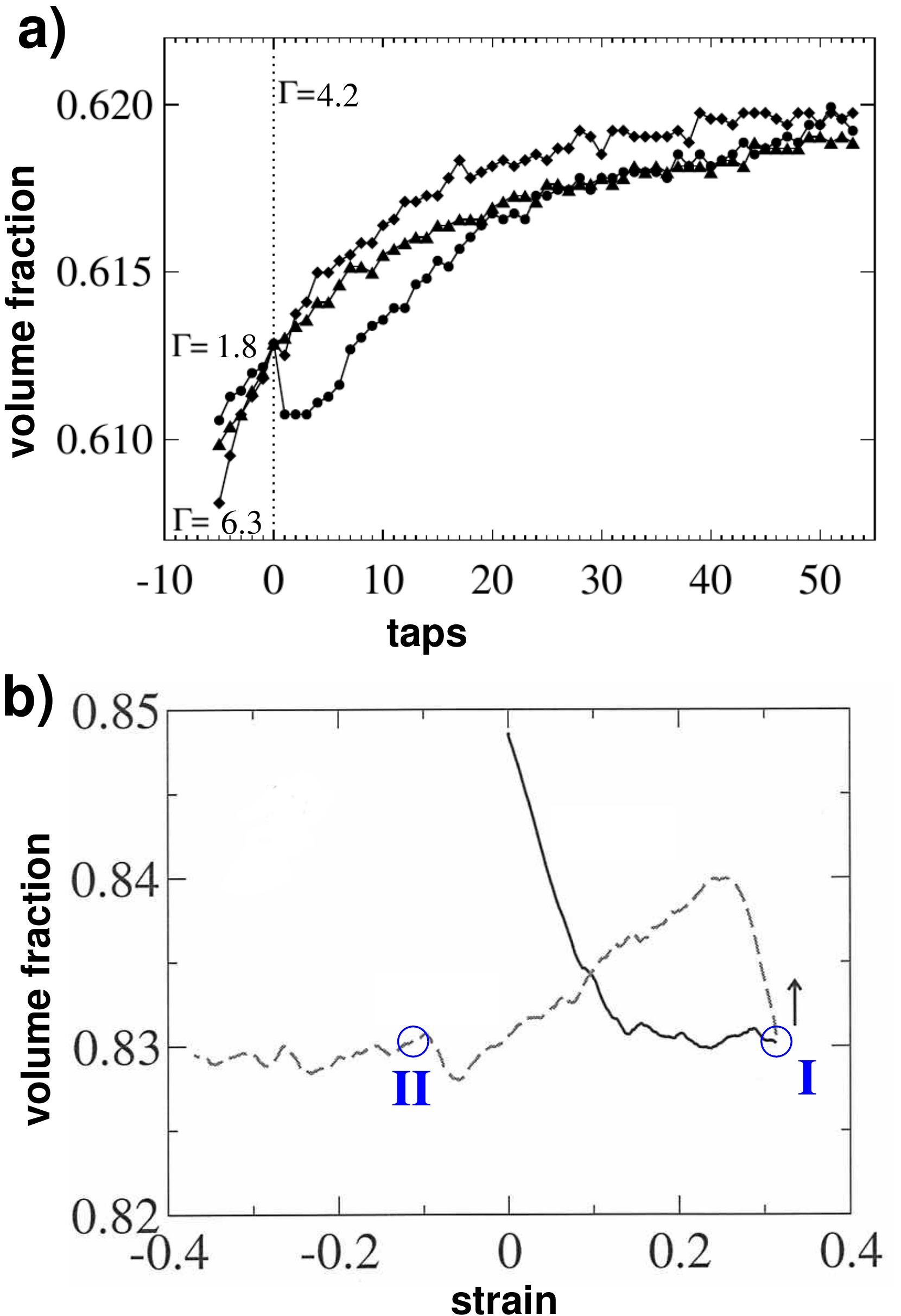}
    \caption{\small
      History dependence in granular systems.
      {\bf a)} Three samples of glass beads are compactified to the same value of $\phi_g$ =0.613  using three different initial 
      tapping strengths. At this point (vertical dotted line)
      the tapping strength is set to the same value of 4.2 g in all three experiments. The system does however respond differently 
      depending on its preparation history. Adapted with permission from \cite{josserand:00}.
     {\bf b)} Volume fraction alone is not sufficient to characterize dilatancy onset. 
      In this contact dynamics simulations an initially dense system of discs has been sheared long enough to dilate to 
      its critical state (solid line). When then the shear direction is reversed (dashed line), the system responds 
      first by compaction before it dilates again. This implies that the packings at points I and II have identical 
      volume fractions but respond differently to shear in the same direction.
      Adapted with permission from \cite{radjai:04}.
    }
    \label{fig:history_dep}
\end{center}
\end{figure}

A number of experiments demonstrate history dependent behavior in granular materials: 
Two seemingly identical packings, which only differ in their preparation histories, respond either differently 
to an external excitation, or they differ in some of their not immediately obvious mechanical properties. 

An example of the first type is shown in figure \ref{fig:history_dep}a: three samples are  
compactified to the same volume fraction $\phi_0$ but using three different driving strengths $\Gamma_i$.  
When these samples at $\phi_0$ are then driven with the same strength $\Gamma_0$, their response depends on the past $\Gamma_i$, 
not $\Gamma_0$ \cite{josserand:00}. Similar results can be obtained for periodically shearing
glass spheres in a parallelepiped shear cell \cite{nicolas:00} or going to large strains in a simple shear cell 
(fig. \ref{fig:history_dep}b) \cite{radjai:04}.

Examples of the second type of history dependence include how the pressure distribution below a sandpile 
depends on its preparation history:
If the sand rained down from a large sieve, the maximum pressure at the bottom plate will be below the tip; 
which is the point with the largest column of sand on top. However, if the sand flowed out of a small funnel opening, 
which means that the pile grew from many downhill avalanches, the maximum pressure at the bottom plate will be at a ring
with a diameter of roughly one third of the total pile diameter \cite{vanel:99_pile}. An similar example is the
history dependence that exists in the so called Janssen effect: The pressure at the bottom of a cylindrical column filled with grains will 
be lower than the total weight of the grains because tangential forces at the sidewalls carry a part of the load. The amount of 
this reduction will again depend on the preparation history \cite{vanel:99}.
Finally it has also been shown numerically, that the number of contacts formed in a packing depends on the preparation history
\cite{agnolin:07}.

History dependence does also exist in frictionless packings (see \cite{kumar:16,luding:16} 
for a novel approach how the jamming volume fraction can be used as a state variable 
to characterize the history). However, most of the examples listed above seem to require friction. 
Either because the extra degrees of freedom allow variability in the contact number or the geometric 
fabric formed by the contacts (pressure distribution at the sand pile bottom, shear response at critical state). 
Or because the memory of a previous state can be encoded as 
a particular configuration in the force phase space spanned by 
hyperstaticity (Janssen).

In all examples discussed here the apparent identity of the initial states has 
only been established in terms of global variables such as shape of the 
sample and $\phi_g$. In fact, in all these cases history dependence 
can also be viewed as another name for:  
"we do not know all relevant parameters which characterize the system".

\section{Granular matter is rarely spherical.}
\label{sec:shape}
\begin{figure}[t]
\begin{center}
  \includegraphics[width=0.40\textwidth]{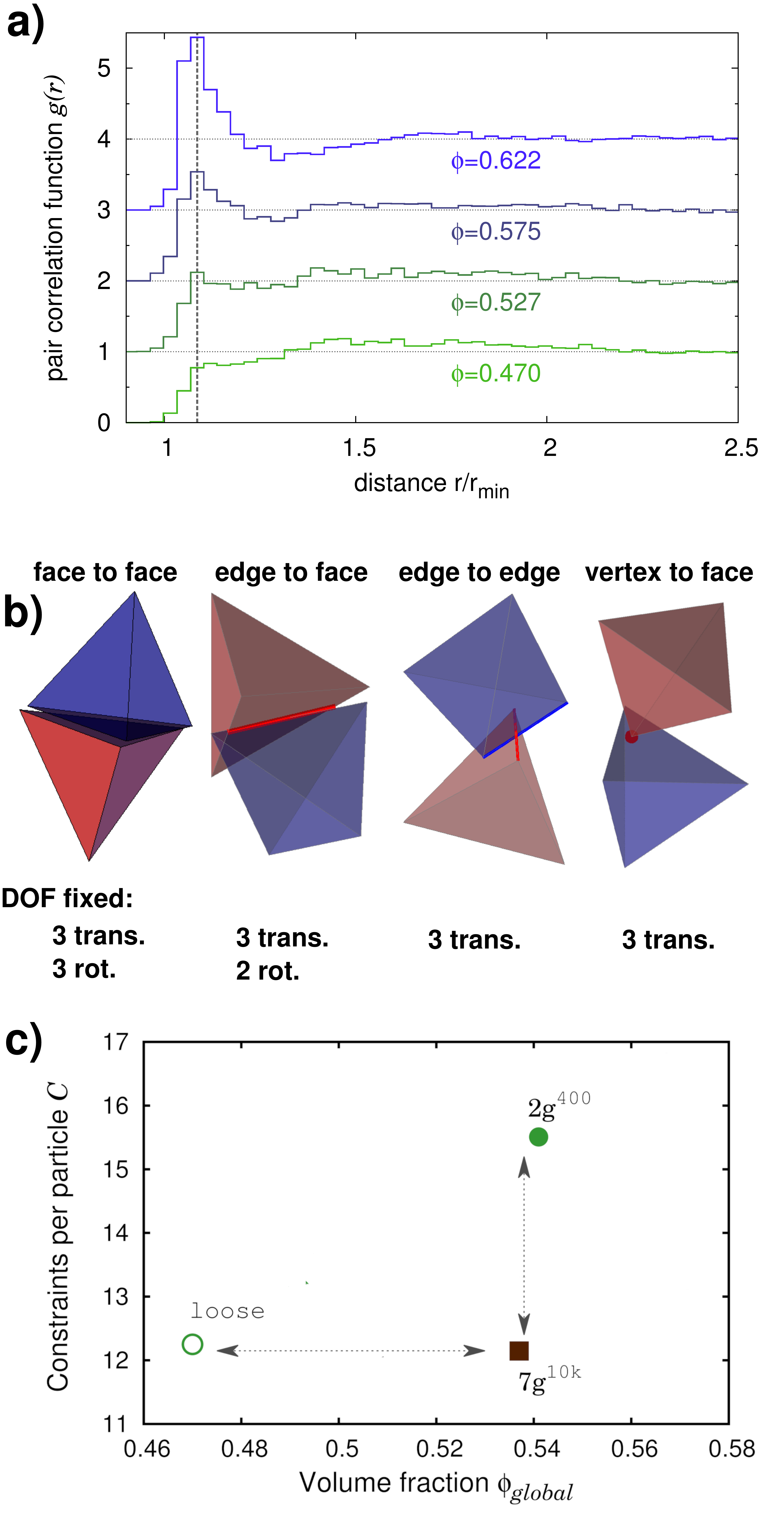}
    \caption{\small
      Packings of tetrahedra differ from sphere packings.
      {\bf a)} The pair correlation functions of experimental tetrahedra packings shows that contrary to spheres the shortest possible 
      distance $r_\textnormal{min}$ (approximately 0.408 times the side-length) is
      not the most likely distance between individual particles (indicated by a vertical dashed line) \cite{neudecker:13}.
      Offsets have been added for improved readability.
      {\bf b)} Tetrahedra form four different types of contacts, which block different numbers of translational and rotational degrees
      of freedom. For the purpose of defining a distance to isostaticity we need to determine the total
      number of constraints per particle $C$ blocked by the combination of all four contact types.
     {\bf c)} Isostaticity corresponds to $C=6$ (each tetrahedra has three rotational and three translational DOF).
     Even the loosest tetrahedra packings have twice the constraint number needed for isostaticity, which means that 
     tetrahedra packings are much more hyperstatic than sphere packings. Moreover do tetrahedra packings exhibit a strong
     history dependence. Using different tapping protocols it is possible to create pairs of packing which differ strongly in only
     one of the two variables $C$ and $\phi_g$ \cite{thyagu:15}.
   }
    \label{fig:tetra}
\end{center}
\end{figure}

Most of the readers will have heard some variant of the "spherical cow in vacuum" joke \cite{spherical_cow}.
But in fact there are not only good reasons for theorists to use spheres as a first approximation, also numerical scientists 
appreciate the easy collision detection algorithm coming with spheres. And experimentalist like spheres because they
are the only monoschematic particles (all particles have the same shape) which are easily available in large quantities.
Being monoschematic is a big advantage during image processing where the a priori knowledge of their shape
helps to identify the individual particles \cite{weis:17}.

Still, real world granular materials are basically always non-spherical in shape. This adds 
additional complexity which every theory suitable for practical purposes will need to take into account.
Figure \ref{fig:tetra} discusses some of this complexity using packings of tetrahedra as an example. 
Contrary to spheres, tetrahedra can form four different types of contacts, cf. figure \ref{fig:tetra}b.
This has important consequences for the pair correlation function shown in figure \ref{fig:tetra}a.
Because  perfectly aligned  face-to-face contacts are less frequent 
than slightly shifted face-to-face or low angle face-to-edge contacts, the closest possible distance between two
particles is {\it not} the most likely contact configuration \cite{neudecker:13}.
The different shape of the first peak of $g(r)$ brings as a consequence that
scaling laws developed for compressible sphere packings \cite{van_hecke:10,liu:10}, such as eq.~\ref{eq:jamming},
will not work for compressible tetrahedra.
Moreover, figure \ref{fig:tetra} c) shows, that tetrahedra packings are even stronger 
hyperstatic and history-dependent than sphere packings.  

Experimentally, one way of moving away from the spherical cow paradigm 
is to take advantage of the quickly improving 3D printing technology to create 
large samples of monoschematic but non-spherical particles \cite{athanassiadis:13,scholz:16}.
Alternatively, we can directly use natural materials such as sand and improve our 3D image processing to obtain
trustworthy segmentation results \cite{vlahinic:14}.
Additionally, there is also progress towards theories for non-spherical particles \cite{baule:13,baule:14}.

\section*{Conclusion}
Frictionless spheres are a great model for emulsions, foams, glasses, and colloids. They give reasonable results for 
granular gases and describe even glassy behavior in driven granular systems. 
But a {\it frictionless granular packing} is a  
a self-contradicting  statement, describing a theoretical model that for the most part has outlived its usefulness.

\section*{Acknowledgments and References}
The ideas in this article have taken shape  in many invaluable discussions with colleagues:
Vasili Baranau, Karen Daniels, Max Neudecker, Charles Radin, Fabian Schaller, Gerd Schr\"oder-Turk, 
Harry Swinney, Brian Tighe, Song-Chuan Zhao.


\begin{thebibliography}{77}

\bibitem{schroeter:05}
M.~Schr\"oter, D.I. Goldman, H.L. Swinney, Phys. Rev. E \textbf{71}, 030301
  (2005)

\bibitem{farrell:10}
G.R. Farrell, K.M. Martini, N.~Menon, Soft Matter \textbf{6}, 2925 (2010)

\bibitem{bowden:73}
F.P. Bowden, D.~Tabor, \emph{Friction} (Anchor Books, Garden City, 1973)

\bibitem{dijksman:13}
J.A. Dijksman, H.~Zheng, R.P. Behringer, in \emph{Powders \& Grains} (2013),
  Vol. 1542 of \emph{AIP Conf. Proc.}, pp. 457--460

\bibitem{dijksman:17}
J.~Dijksman, \emph{private communication} (2017)

\bibitem{gao:09}
G.J. Gao, J.~Blawzdziewicz, C.S. O'Hern, M.~Shattuck, Phys. Rev. E \textbf{80},
  061304 (2009)

\bibitem{van_hecke:10}
M.~{van Hecke}, J. Phys.: Condens. Matter \textbf{22}, 033101 (2010)

\bibitem{tighe:10}
B.P. Tighe, J.H. Snoeijer, T.J.H. Vlugt, M.v. Hecke, Soft Matter \textbf{6},
  2908 (2010)

\bibitem{shundyak:07}
K.~Shundyak, M.~van Hecke, W.~van Saarloos, Phys. Rev. E \textbf{75}, 010301
  (2007)

\bibitem{silbert:02}
L.E. Silbert, D.~Erta\c{s}, G.S. Grest, T.C. Halsey, D.~Levine, Phys. Rev. E
  \textbf{65}, 031304 (2002)

\bibitem{zhang:05}
H.P. Zhang, H.A. Makse, Phys. Rev. E \textbf{72}, 011301 (2005)

\bibitem{henkes:10}
S.~Henkes, M.~{van Hecke}, W.~{van Saarloos}, Europhys. Lett. \textbf{90},
  14003 (2010)

\bibitem{edwards:89}
S.F. Edwards, R.B.S. Oakeshott, Physica A \textbf{157}, 1080 (1989)

\bibitem{mehta:89}
A.~Mehta, S.F. Edwards, Physica A \textbf{157}, 1091 (1989)

\bibitem{anikeenko:07}
A.V. Anikeenko, N.N. Medvedev, Phys. Rev. Lett. \textbf{98}, 235504 (2007)

\bibitem{radin:08}
C.~Radin, J. Stat. Phys. \textbf{131}, 567 (2008)

\bibitem{jin:10}
Y.~Jin, H.A. Makse, Physica A \textbf{389}, 5362 (2010)

\bibitem{kapfer:12}
S.C. Kapfer, W.~Mickel, K.~Mecke, G.E. Schr\"oder-Turk, Phys. Rev. E
  \textbf{85}, 030301 (2012)

\bibitem{francois:13}
N.~Francois, M.~Saadatfar, R.~Cruikshank, A.~Sheppard, Phys. Rev. Lett.
  \textbf{111}, 148001 (2013)

\bibitem{baranau_random_close:14}
V.~Baranau, U.~Tallarek, Soft Matter \textbf{10}, 3826 (2014)

\bibitem{baranau_jamming:14}
V.~Baranau, U.~Tallarek, Soft Matter \textbf{10}, 7838 (2014)

\bibitem{torquato:07}
S.~Torquato, F.H. Stillinger, J. App. Phys. \textbf{102}, 093511 (2007)

\bibitem{parisi:10}
G.~Parisi, F.~Zamponi, Rev. Mod. Phys. \textbf{82}, 789 (2010)

\bibitem{asenjo:14}
D.~Asenjo, F.~Paillusson, D.~Frenkel, Phys. Rev. Lett. \textbf{112}, 098002
  (2014)

\bibitem{martiniani:16}
S.~Martiniani, K.J. Schrenk, J.D. Stevenson, D.J. Wales, D.~Frenkel, Phys. Rev.
  E \textbf{93}, 012906 (2016)

\bibitem{ohern:03}
C.S. O'Hern, L.E. Silbert, A.J. Liu, S.R. Nagel, Phys. Rev. E \textbf{68},
  011306 (2003)

\bibitem{pica_ciamarra:10}
M.~Pica~Ciamarra, M.~Nicodemi, A.~Coniglio, Soft Matter \textbf{6}, 2871 (2010)

\bibitem{kumar:16}
N.~Kumar, S.~Luding, Granul. Matter \textbf{18}, 58 (2016)

\bibitem{onoda:90}
G.Y. Onoda, E.G. Liniger, Phys. Rev. Lett. \textbf{64}, 2727 (1990)

\bibitem{jerkins:08}
M.~Jerkins, M.~Schr\"oter, H.L. Swinney, T.J. Senden, M.~Saadatfar, T.~Aste,
  Phys. Rev. Lett. \textbf{101}, 018301 (2008)

\bibitem{song:08}
C.~Song, P.~Wang, H.A. Makse, Nature \textbf{453}, 629 (2008)

\bibitem{silbert:10}
L.E. Silbert, Soft Matter \textbf{6}, 2918 (2010)

\bibitem{delaney:11}
G.W. Delaney, J.E. Hilton, P.W. Cleary, Phys. Rev. E \textbf{83}, 051305 (2011)

\bibitem{srebro:03}
Y.~Srebro, D.~Levine, Phys. Rev. E \textbf{68}, 061301 (2003)

\bibitem{baranau:16}
V.~Baranau, S.C. Zhao, M.~Scheel, U.~Tallarek, M.~Schr\"oter, Soft Matter
  \textbf{12}, 3991 (2016)

\bibitem{briscoe:08}
C.~Briscoe, C.~Song, P.~Wang, H.A. Makse, Phys. Rev. Lett. \textbf{101}, 188001
  (2008)

\bibitem{mcnamara:09}
S.~McNamara, P.~Richard, S.K. de~Richter, G.~Le~Caer, R.~Delannay, Phys. Rev. E
  \textbf{80}, 031301 (2009)

\bibitem{zhao:14}
S.C. Zhao, M.~Schr\"oter, Soft Matter \textbf{10}, 4208 (2014)

\bibitem{nowak:98}
E.R. Nowak, J.B. Knight, E.~Ben-Naim, H.M. Jaeger, S.R. Nagel, Phys. Rev. E
  \textbf{57}, 1971 (1998)

\bibitem{ribiere:07}
P.~Ribi\`ere, P.~Richard, P.~Philippe, D.~Bideau, R.~Delannay, Eur. Phys. J. E
  \textbf{22}, 249 (2007)

\bibitem{andreotti:13}
B.~Andreotti, Y.~Forterre, O.~Pouliquen, \emph{Granular Media} (Cambridge
  University Press, 2013)

\bibitem{schroeter:07}
M.~Schr\"oter, S.~N\"agle, C.~Radin, H.L. Swinney, Europhys. Lett. \textbf{78},
  44004 (2007)

\bibitem{kabla:09}
A.J. Kabla, T.J. Senden, Phys. Rev. Lett. \textbf{102}, 228301 (2009)

\bibitem{gravish:10}
N.~Gravish, P.B. Umbanhowar, D.I. Goldman, Phys. Rev. Lett. \textbf{105},
  128301 (2010)

\bibitem{umbanhowar:10}
P.~Umbanhowar, D.I. Goldman, Phys. Rev. E \textbf{82}, 010301 (2010)

\bibitem{metayer:11}
J.~M\'etayer, D.J. Suntrup~{III}, C.~Radin, H.L. Swinney, M.~Schr\"oter,
  Europhys. Lett. \textbf{93}, 64003 (2011)

\bibitem{mueggenburg:12}
N.~Mueggenburg, Phys. Rev. E \textbf{85}, 041305 (2012)

\bibitem{bi:11}
D.~Bi, J.~Zhang, B.~Chakraborty, R.P. Behringer, Nature \textbf{480}, 355
  (2011)

\bibitem{ren:13}
J.~Ren, J.A. Dijksman, R.P. Behringer, Phys. Rev. Lett. \textbf{110}, 018302
  (2013)

\bibitem{vinutha:16}
H.A. Vinutha, S.~Sastry, Nature Phys. \textbf{12}, 578 (2016)

\bibitem{luding:16}
S.~Luding, Nature Phys. \textbf{12}, 531 (2016)

\bibitem{katgert:10}
G.~Katgert, M.~{van Hecke}, Europhys. Lett. \textbf{92}, 34002 (2010)

\bibitem{utermann:11}
S.~Utermann, P.~Aurin, M.~Benderoth, C.~Fischer, M.~Schr\"oter, Phys. Rev. E
  \textbf{84}, 031306 (2011)

\bibitem{majmudar:07}
T.S. Majmudar, M.~Sperl, S.~Luding, R.P. Behringer, Phys. Rev. Lett.
  \textbf{98}, 058001 (2007)

\bibitem{coulais:14}
C.~Coulais, R.P. Behringer, O.~Dauchot, Soft Matter \textbf{10}, 1519 (2014)

\bibitem{lechenault:06}
F.~Lechenault, F.d. Cruz, O.~Dauchot, E.~Bertin, J. Stat. Mech. \textbf{2006},
  P07009 (2006)

\bibitem{zhao:12}
S.C. Zhao, S.~Sidle, H.L. Swinney, M.~Schr\"oter, Europhys. Lett. \textbf{97},
  34004 (2012)

\bibitem{aste:06}
T.~Aste, M.~Saadatfar, T.J. Senden, J. Stat. Mech.: Theory and Exp.
  \textbf{2006}, P07010 (2006)

\bibitem{baule:13}
A.~Baule, R.~Mari, L.~Bo, L.~Portal, H.A. Makse, Nat. Commun. \textbf{4}, 2194
  (2013)

\bibitem{baule:14}
A.~Baule, H.A. Makse, Soft Matter \textbf{10}, 4423 (2014)

\bibitem{schaller:15}
F.M. Schaller, M.~Neudecker, M.~Saadatfar, G.W. Delaney, G.E. Schr\"oder-Turk,
  M.~Schr\"oter, Phys. Rev. Lett. \textbf{114}, 158001 (2015)

\bibitem{schroeder-turk:10}
G.E. {Schr\"oder-Turk}, W.~Mickel, M.~Schr\"oter, G.W. Delaney, M.~Saadatfar,
  T.J. Senden, K.~Mecke, T.~Aste, Europhys. Lett. \textbf{90}, 34001 (2010)

\bibitem{radjai:09}
F.~Radjai, in \emph{Powders \& Grains} (2009), Vol. 1145 of \emph{AIP Conf.
  Proc.}, pp. 35--42

\bibitem{josserand:00}
C.~Josserand, A.V. Tkachenko, D.M. Mueth, H.M. Jaeger, Phys. Rev. Lett.
  \textbf{85}, 3632 (2000)

\bibitem{radjai:04}
F.~Radja\"i, S.~Roux, in \emph{The Physics of Granular Media}, edited by
  H.~Hinrichsen, D.E. Wolf (Wiley-VCH, Weinheim, 2004), pp. 165--187

\bibitem{nicolas:00}
M.~Nicolas, P.~Duru, O.~Pouliquen, Eur. Phys. J. E \textbf{3}, 309 (2000)

\bibitem{vanel:99_pile}
L.~Vanel, D.~Howell, D.~Clark, R.P. Behringer, E.~Cl\'ement, Phys. Rev. E
  \textbf{60}, R5040 (1999)

\bibitem{vanel:99}
L.~Vanel, E.~Cl\'ement, Eur. Phys. J. B \textbf{11}, 525 (1999)

\bibitem{agnolin:07}
I.~Agnolin, J.N. Roux, Phys. Rev. E \textbf{76}, 061302 (2007)

\bibitem{neudecker:13}
M.~Neudecker, S.~Ulrich, S.~Herminghaus, M.~Schr\"oter, Phys. Rev. Lett.
  \textbf{111}, 028001 (2013)

\bibitem{thyagu:15}
N.N. Thyagu, F.~Schaller, S.~Weis, M.~Neudecker, M.~Schr\"oter,
  arXiv:1501.04472  (2015)

\bibitem{spherical_cow}
\url{https://www.youtube.com/watch?v=oUwlEdz42xo}

\bibitem{weis:17}
S.~Weis, M.~Schr\"oter, submitted to {Rev. Sci. Instr.}  (2017),
  http://arxiv.org/abs/arXiv:1612.06639

\bibitem{liu:10}
A.J. Liu, S.R. Nagel, Ann. Rev. Cond. Matt. Phys. \textbf{1}, 347 (2010)

\bibitem{athanassiadis:13}
A.G. Athanassiadis, M.Z. Miskin, P.~Kaplan, N.~Rodenberg, S.H. Lee, J.~Merritt,
  E.~Brown, J.~Amend, H.~Lipson, H.M. Jaeger, Soft Matter \textbf{10}, 48
  (2013)

\bibitem{scholz:16}
C.~Scholz, S.~D’Silva, T.~P\"oschel, New J. Phys. \textbf{18}, 123001 (2016)

\bibitem{vlahinic:14}
I.~Vlahini\'c, E.~And\`o, G.~Viggiani, J.E. Andrade, Granul. Matter
  \textbf{16}, 9 (2014)

\end{thebibliography}
\end{document}